\documentclass[prl,11pt,preprint]{revtex4-1}

\usepackage{amsmath}    
\usepackage{graphicx}   
\usepackage{verbatim}   
\usepackage{color}      
\usepackage{subfigure}  
\usepackage{hyperref}   
\begin{document}

\title{Hassle-free Approach to Thermal Transport Measurements Using Spatial-Temporal Temperature Data }
\author{Ding Ding and Kedar Hippalgaonkar}
\affiliation{Institute of Materials Research and Engineering, Agency for Science Technology and Research, 2 Fusionopolis Way, Singapore 138634}

\date{\today}

\begin{abstract}
Nanoscale engineering and novel materials have created interesting effects in thermal transport. Thermal conductivity can now be different due to physical and heating sizes. Also, highly anisotropic thermal conductivity can result from unique material composition and geometries. Various experimental methods have been developed to measure these thermal conductivity variations. All of them require varying the physical size of the sample, the heating size or relative positions between heating and detection. Here, we numerically propose a time-domain optical method that uses spatial temporal temperature data to resolve anisotropic and size-dependent thermal conductivity. Our method is hassle-free as it does not vary any experimental parameters and is easily compatible with various methods of measuring temperature in the time domain. This technique can high throughput screening of thermal properties for nanoengineered and novel materials in thermal transport. Also, this technique can be used to identify novel effects in thermal transport within a single experiment.

\end{abstract}

\maketitle

\section{Introduction}

A good knowledge of the material's thermal conductivity is essential to managing heat flow in all thermal design and analysis. In most materials, their thermal conductivity is isotropic or direction independent. Recently, novel materials such as layered van der Waals solids \cite{slack_anisotropic_1962,liu_measurement_2014,lee_anisotropic_2015,luo_anisotropic_2015,lindroth_thermal_2016} and superlattices \cite{luckyanova_anisotropy_2013} have been shown to have anisotropic thermal conductivity. This allows for applications such as heat spreaders which conducts heat away very fast in one direction but limits heat flow in another direction \cite{suszko_thermally_2016}. Due to the anisotropy, experimental measurements developed so far generally requires multiple measurements along different crystal axes \cite{feser_probing_2012} . Other methods require variations of heating size\cite{schmidt_pulse_2008,liu_simultaneous_2013,jiang_time-domain_2017} or anisotropic heating or anisotropic detection in order to change or detect the desired direction of heat transport \cite{liu_measurement_2014,bogner_cross-_2017,li_anisotropic_2018}. Of course, independent measurements can be performed along different directions of the same anisotropic material \cite{luckyanova_anisotropy_2013,lee_anisotropic_2015,kim_elastic_2017} to obtain the respective thermal conductivities.

Another interesting effect that is typically observed is size-dependent thermal conductivity. This happen when the experimentally measured thermal conductivity is lower than the bulk value. Experiments variations of sample size \cite{asheghi_phonon-boundary_1997,li_thermal_2003,hsiao_observation_2013,zhang_temperature_2015,ramu_electrical_2016} and heating length scales in both time \cite{siemens_quasi-ballistic_2010,minnich_thermal_2011,johnson_direct_2013,hu_spectral_2015} and frequency domain \cite{koh_frequency_2007,regner_broadband_2013} have both observed such effects. Electrical measurement methods and optical methods that directly vary heater sizes generally require multiple samples to be fabricated. Optical techniques that uses the same sample also require numerous spatial or temporal variations of the heating beam in order to observe such effects.

Here, we propose an experimental method that uses spatial temporal temperature data to retrieve multiple thermal transport parameters in a single experiment. First, we demonstrate numerically that multiple thermal conductivity values in a multilayers system can be retrieved once sufficient points have been sampled. Second, we show that size-dependent thermal conductivity variations for an anisotropic material can be accurately recovered. Last but not least, we discuss the compatibility of our method to current measurement techniques and conclude possibility of extending our work to other thermal effects of interest.

\section{Methodology}

\begin{figure}[h!]
    \centering
        \includegraphics[width=0.45\textwidth]{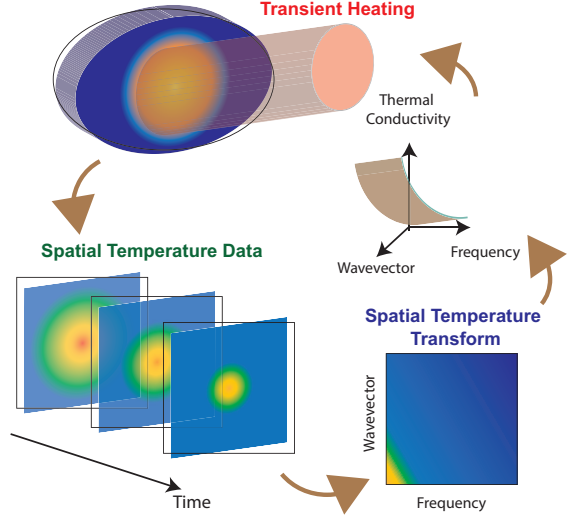}
     ~
            \caption{ Schematic of our proposed method to obtain multiple thermal transport parameters. Cylindrical symmetry is assumed in this setup. } \label{fig:aniso_case}
\end{figure}

Here, we assume a gaussian shaped transient heat flux on the sample surface at $t=0$ as shown in Fig. \ref{fig:aniso_case}. This simulates gaussian heating on a sample of finite thickness. The boundary condition bottom surface is assumed to be adiabatic. The proposed method then captures temperature versus time data on the material's surface and performing a spatial temporal transformation. The data in the transformed space is represented as a function of the spatial wavevector and the temporal frequency. Then, thermal conductivity parameters can be solved all at once using data from the transformed space. The temperature profile on the top surface changes with time according to Fourier's Law assuming radial symmetry in cylindrical coordinates, the heat equation is written as \cite{cahill_analysis_2004} $\frac{k_r}{r}\frac{\partial}{\partial r} (r \frac{\partial \Delta T}{\partial r}) +k_z \frac{\partial^2  \Delta T}{\partial z^2}=\rho c_p \frac{\partial  \Delta T}{\partial t}$ where $k_r$ and $k_z$ are the radial and cross-plane thermal conductivities, respectively. $\rho$ is the density of the material and $c_p$ is the specific heat. We can be Hankel and then Fourier transformed to yield

\begin{equation}
\frac{\partial^2 {\bar{ \Delta T}(\kappa,\omega)}}{\partial z^2}=q^2 {\bar{ \Delta T}(\kappa,\omega)}
\label{eq:hankel_fourier}
\end{equation}
 
 where $q^2(\kappa,\omega)=\frac{k_r \kappa^2+\rho c i \omega}{k_z}$. $\kappa$ and $\omega$ are the Hankel spatial wavevector and Fourier transformed frequency respectively.  
 The radial definition in Eq. \ref{eq:hankel_fourier} can be solved for geometries containing more than one layer \cite{carslaw_conduction_1986,feldman_algorithm_1999}. The matrix describing any of the layers can be written as 
 
 \begin{equation}
 \begin{bmatrix}
           \bar{ \Delta T}_b\\
           f_b
         \end{bmatrix}=
         \begin{pmatrix}
         A & B \\
C & D \hfill 
\end{pmatrix}
 \begin{bmatrix}
           \bar{ \Delta T}_t\\
           f_t
         \end{bmatrix}
\label{eq:matrix}
\end{equation}
 
where $\bar{ \Delta T}_b$ and $\bar{ \Delta T}_t$ are the back and top temperatures in the transformed space and $f_b$ and $f_t$ are the back and top heat flux. The components are $A=\cosh(q d), B=-\sinh(qd)/(k_z q), C=-k_z q \sinh(qd), D=A $ for a single layer case \cite{carslaw_conduction_1986} where $d$ is the thickness of the layer. For multilayers, each material's matrix in Eq. \ref{eq:matrix} are multiplied together with matrices in between for each interface. For our assumed boundary condition in Fig. \ref{fig:aniso_case}(a) or if the last layer is assume to be semi-infinite, then $\theta_t=-\frac{D}{C} f_t$. 

\section{Anisotropic Thermal Transport}

Figure \ref{fig:aniso_case}(b) typically provides an example of spatially averaged temperature as a function of time after solving Eqs. \ref{eq:hankel_fourier} to \ref{eq:matrix}. The example here is for a multilayer sample shown in Fig. \ref{fig:aniso_case}(b) inset where the top surface is an isotropic material with thermal conductivity $k_{r1}=k_{z1}=205$ W/mK. The bottom material is thermally anisotropic with thermal conductivity $k_{r2}=149.35$ W/mK in-plane and $k_{z2}=1.4935$ W/mK cross-plane. Van der Waal solids such as MoS2 are known to have similar anisotropic thermal conductivities \cite{liu_measurement_2014}. The interface conductance is $G_{12}=120 $MW/m$^2$K between the two materials. This is an example of a typical multilayer sample used when optical reflectance of the top layer provides the temperature information \cite{jiang_tutorial:_2018}. Typically, the top layer is an isotropic material which is relatively thin ($d=100$ nm), so that bulk of the thermal transport happens in the bottom material which is of interest. 

\begin{figure*}[h!]
                   \centering
    \begin{subfigure}[]{
        \includegraphics[width=0.3\textwidth]{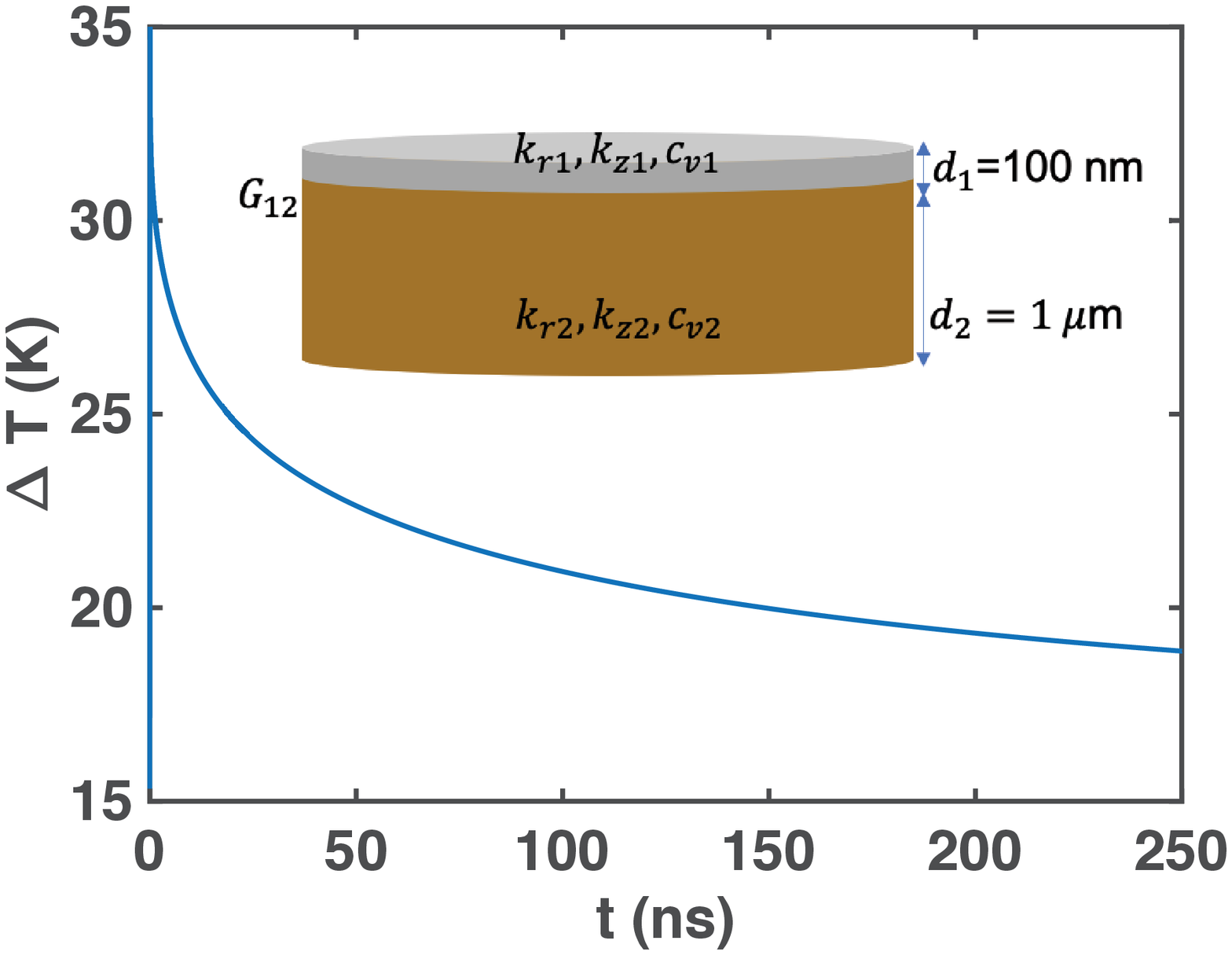}}
    \end{subfigure}
~
          \centering
    \begin{subfigure}[]{
        \includegraphics[width=0.3\textwidth]{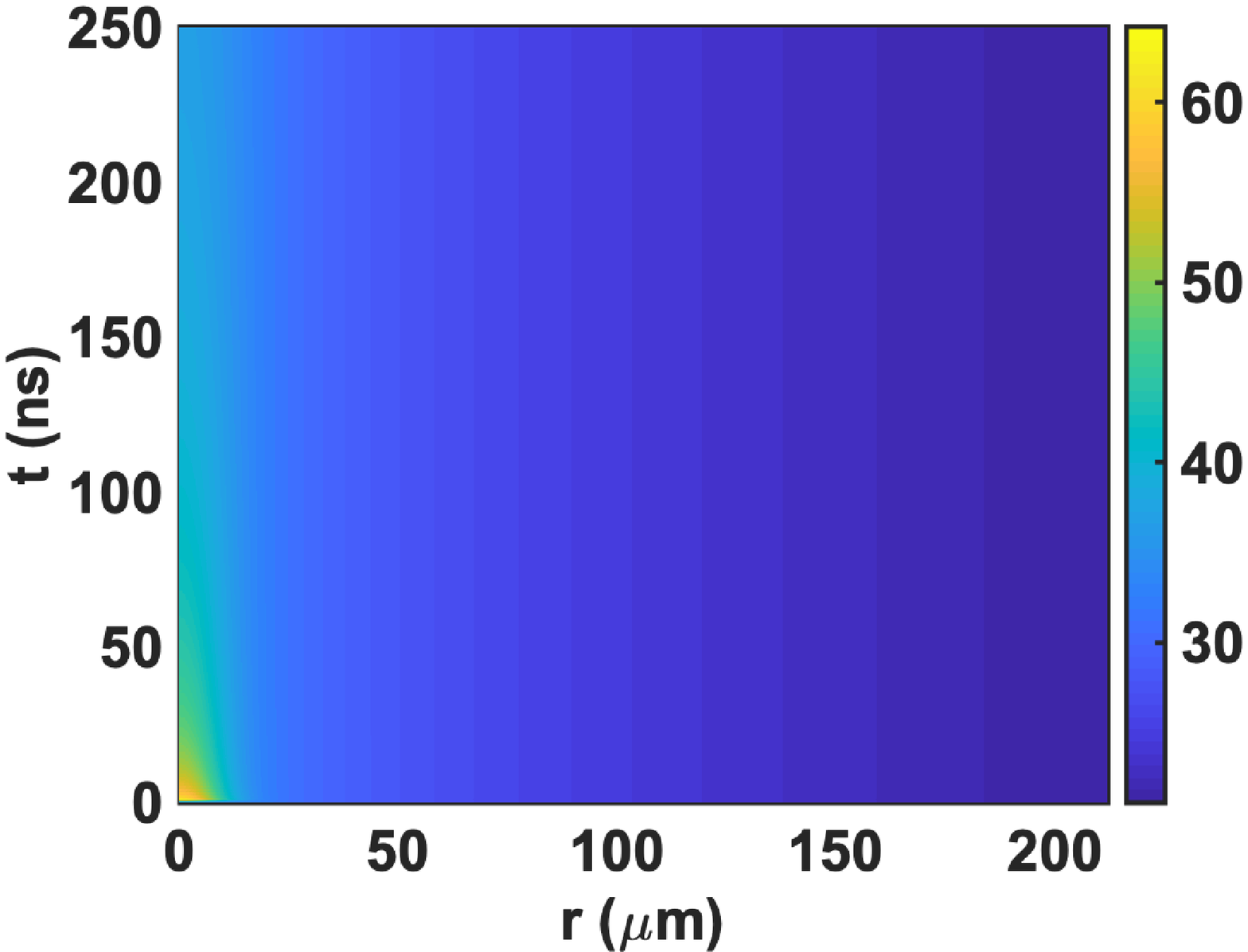}}
    \end{subfigure}
    ~
              \centering
    \begin{subfigure}[]{
        \includegraphics[width=0.3\textwidth]{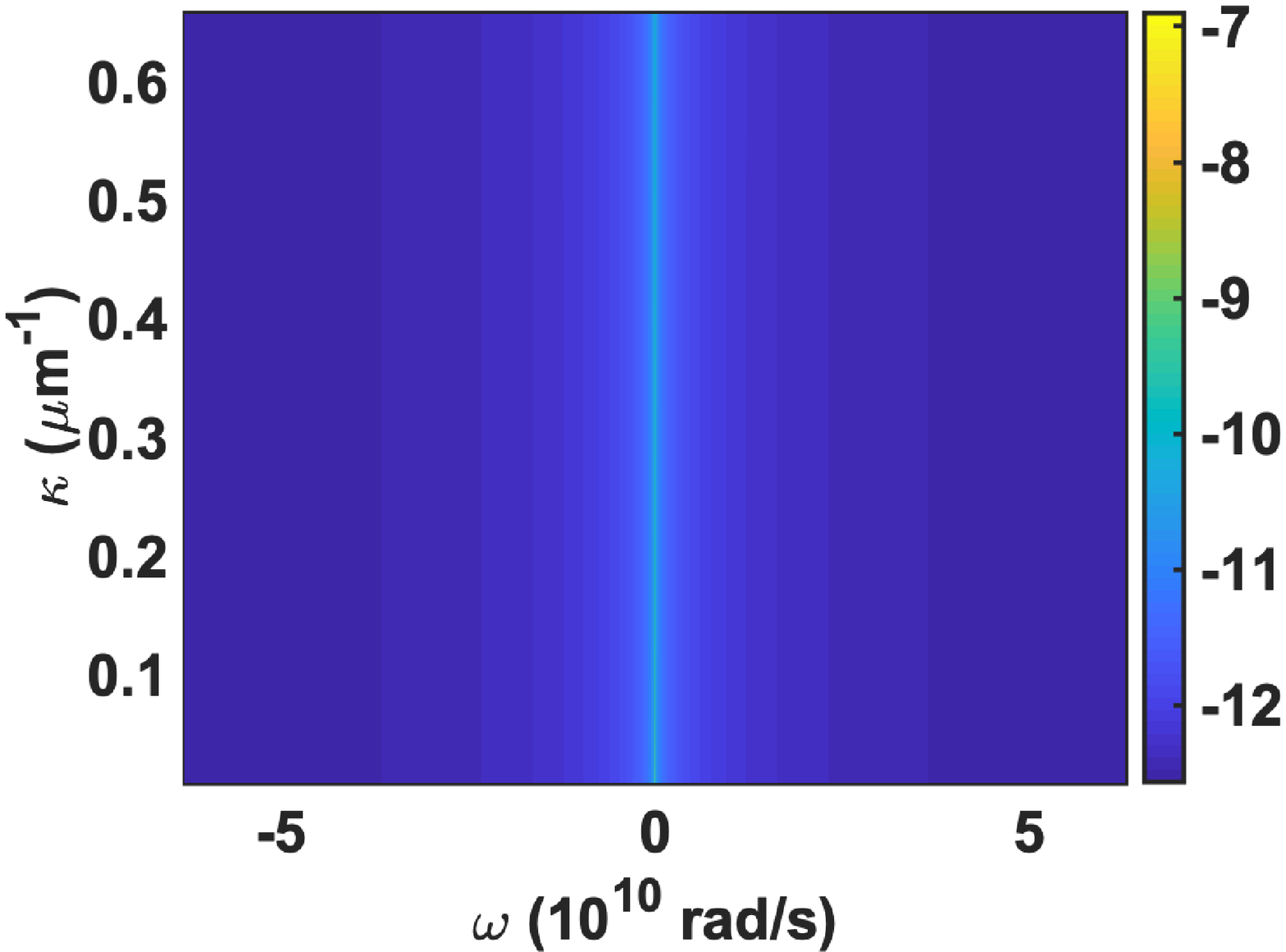}}
    \end{subfigure}
       \caption{  (b) Spatially-averaged temperature as a function of time. The averaging is performed assuming a gaussian profile with same radius as the pump beam. Such spatially-weighted temperature-time plots are typically what is obtained in optical methods to measure thermal transport.  Inset: Schematic of multilayer sample. The top surface is an isotropic material with $k_{r1}=k_{z1}=205$ W/mK. The bottom material is anisotropic with $k_{r2}=149.35$ and $k_{z2}=1.4935$. The interface conductance is $G_{12}=120 $MW/m$^2$K. (a) Temperature profile $\Delta T(r,t)$ for sample as a function of radial distance and time. The heating beam at time $t=0$ is a gaussian beam with radius $w_0=15\mu$m. (b) Transformed $\bar{\Delta T}(\kappa,\omega)$ from the temperature data in (c) using Eq. \ref{eq:hankel_fourier}. The transformed data is plotted in log scale of the absolute value as the values are typically complex. Complex values from (d) can then be used to set up a system of equations for different values of $(\kappa,\omega)$ in order to solve for material parameters in a multiplayer such as in (a).}\label{fig:aniso_temp}
\end{figure*}

Now, we assume that the optically reflective or absorptive top-layer can provide us spatial temporal temperature data. Figure \ref{fig:aniso_temp} (a) shows the temperature profile for $\Delta T(r,t)$ on the sample as a function of time. It can be seen that the thermal decay happens very fast for the first few nanoseconds followed by a slower decay which happens over a longer period of time. If we take a Fourier and Hankel transform using Eq. \ref{eq:hankel_fourier} of the temperature data $\Delta T(r,t)$ in Fig.  \ref{fig:aniso_temp} (a), we can obtain the the transformed $\bar{\Delta T}(\kappa,\omega)$ plotted in Fig. \ref{fig:aniso_temp}(b). Here, if we assume that $k_{r1,2},k_{z1,2}$ and $G_{12}$ are all unknown, we solve for these unknowns from $\bar{\Delta T}(\kappa,\omega)$. For a given set of points in the transformed space $(\kappa,\omega)$ in Fig.  \ref{fig:aniso_temp}(b), each of them is a solution to Eq. \ref{eq:matrix} for the same values of $k_{r1,2},k_{z1,2}$ and $G_{12}$. Doing so for all points in the transformed space allows us to solve a system of simultaneous equations for the values of the transport coefficients $k_{r1,2},k_{z1,2},G_{12}$. Here we do not impose any assumption that $k_{r1}=k_{z1}$ and let the system of equations reveal if each material layer is isotropic or not. The accuracy of the results will depend on the number of points available for  $(\kappa,\omega)$. We would like to highlight that the system of equations is highly non-linear for the multilayered version of Eq. \ref{eq:matrix}. The values of the system of equations are complex and differs by orders of magnitudes as shown in Fig. \ref{fig:aniso_temp}(b). Furthermore, thermal conductivities $k_{r,z}$ are orders of magnitude apart compared to interfacial conductance $G_{12}$, making the system a challenging one to solve in a stable manner. Thus, despite only five unknowns solve, many more equations are generally needed.

\begin{table*}[]
\begin{tabular}{lllllllllll}
\toprule
	 $N_{\omega}$ & $k_{r1}$ & S.E. &	$k_{z1}$ &  S.E. & 	$1/G_{12} $ & S.E. &	$k_{r2} $ &	S.E.  &	$k_{z2}$ & 	S.E.	\\
\hline
 	50	&205	&1.25E-17&	205&	1.42E-20	&8.33E-09	&8.10E-23&	149.35&	1.19E-16	&1.4935	&7.39E-16\\
	25&	205&	7.06E-15	&205	&7.86E-18&	8.33E-09	&4.58E-20	&149.35&	6.73E-14&	1.4935	&4.17E-13\\
	10&	205&	7.81E-15&	-158.92&	8.64E-18	&1.31E-08&	5.07E-20	&148.94	&7.44E-14&	1.5095&	4.62E-13\\
	5	&168.48&	2.05E-06&	197&	1.33E-09&	-8.13E-09	&1.10E-11&	153&	2.02E-05	&1.4576	&0.00010074 \\ \hline\hline
\end{tabular}
\caption{Table of results obtained by varying the number of point $N_{\kappa,\omega}$ extracted from the transformed space $(\kappa,\omega)$. $N_{\kappa}=10$ for all cases shown. S.E. represents the standard error of the mean. A small set of values $N_{\omega}=25$ is sufficient to retrieve actual values accurately.  } \label{table:results} 
\end{table*}

In Table \ref{table:results}, we solved the system of equation using standard non-linear regression methods . We fixed the number of points in $\kappa$ space to be $N_{\kappa}=10$ so as to correspond to the smallest possible spatial resolution in the micrometer range, achievable with conventional imaging methods. Then, we vary the number of points in $\omega$ with $\omega_{max}=1\times 10^{10}$ rad/s. As we increase $N_\omega$, the observed experiment time increases and the solutions become more accurate. As shown As little as $N_\omega=25$ points is sufficient to reconstruct all values accurately. 
   
\section{Anisotropy with Size Dependence}

\begin{figure}[h!]
    \centering
    \begin{subfigure}[]{
        \includegraphics[scale=0.4]{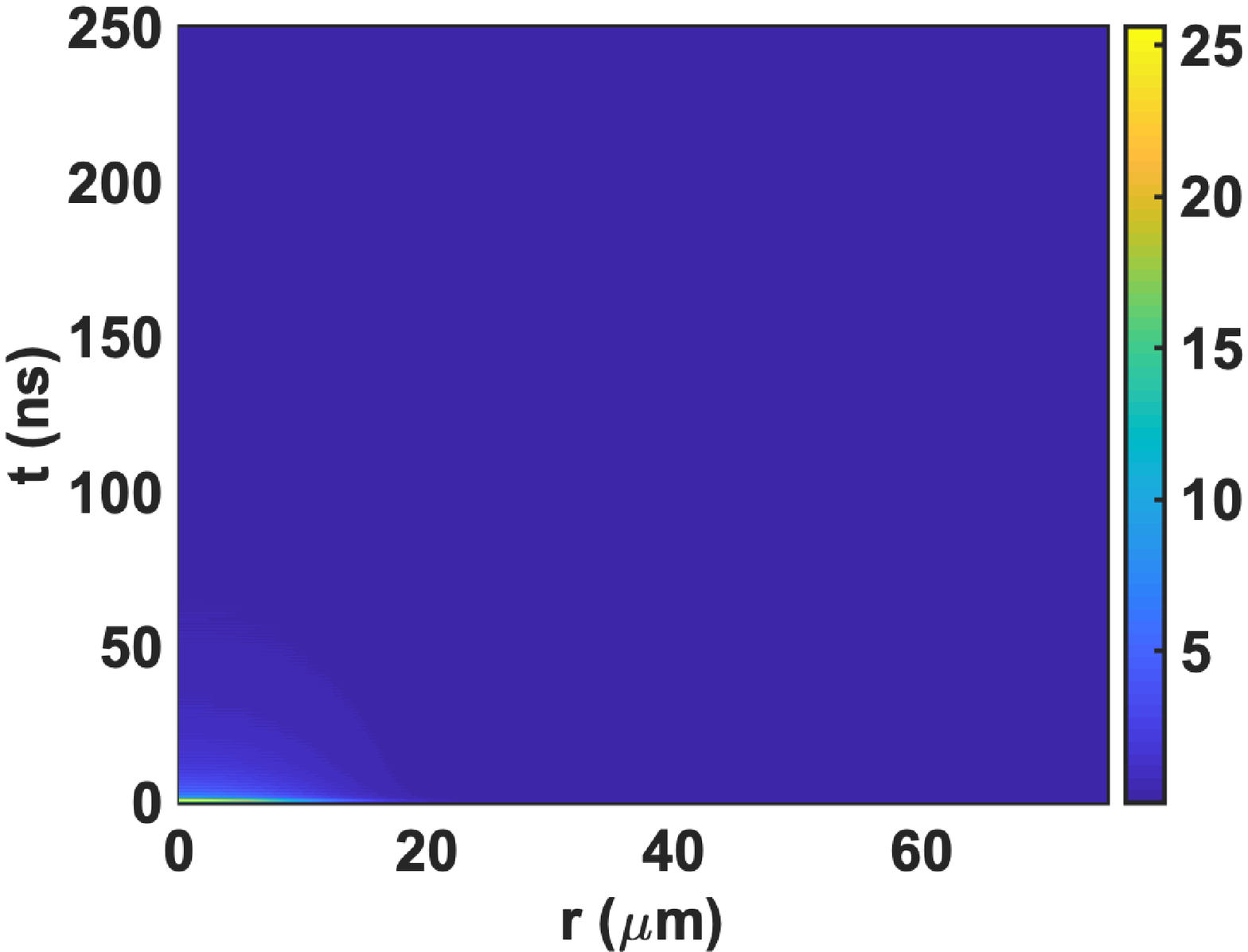}}
    \end{subfigure}
          \centering
    \begin{subfigure}[]{
        \includegraphics[scale=0.4]{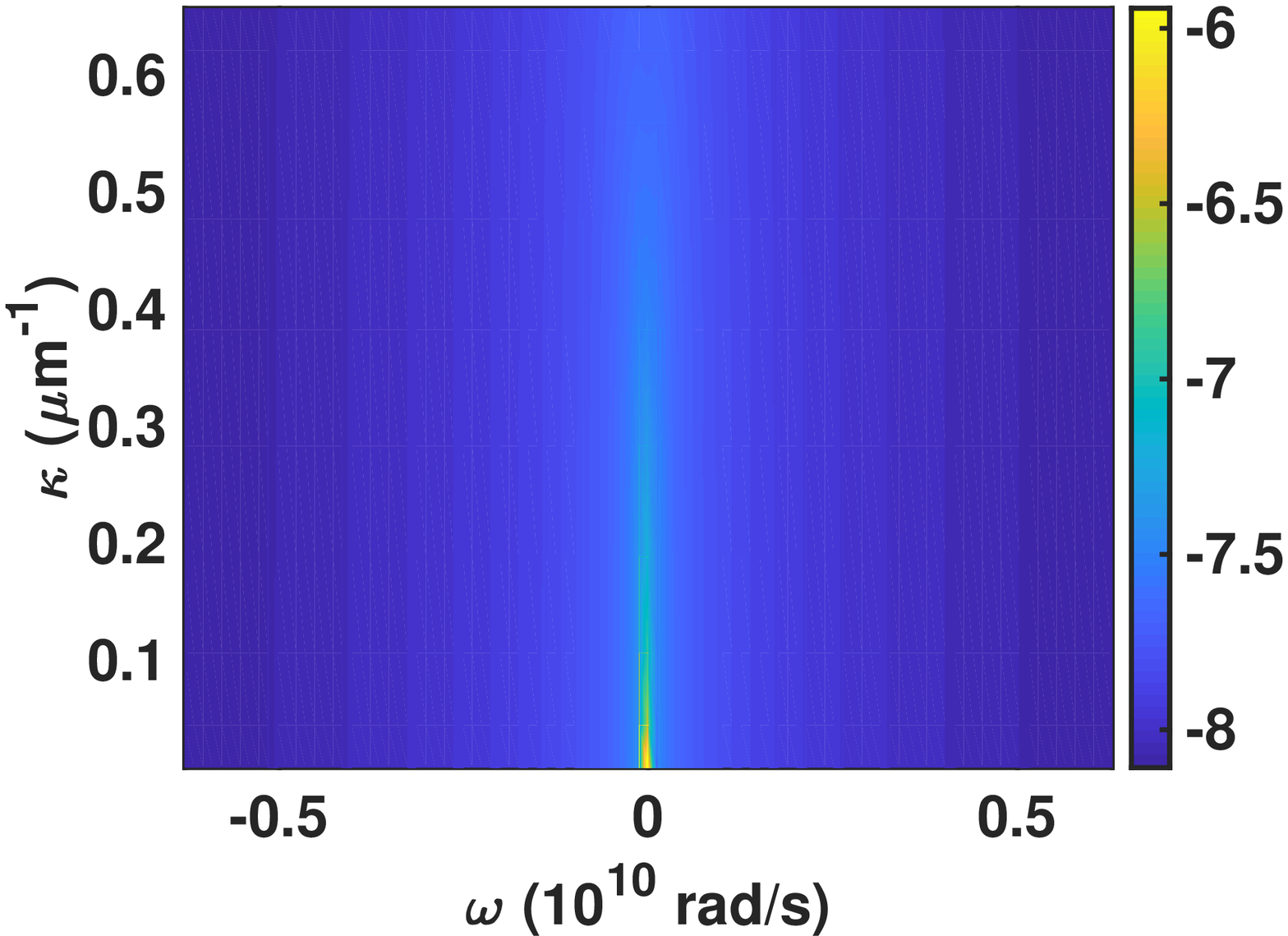}}
    \end{subfigure}  
              \centering
    \begin{subfigure}[]{
        \includegraphics[scale=0.4]{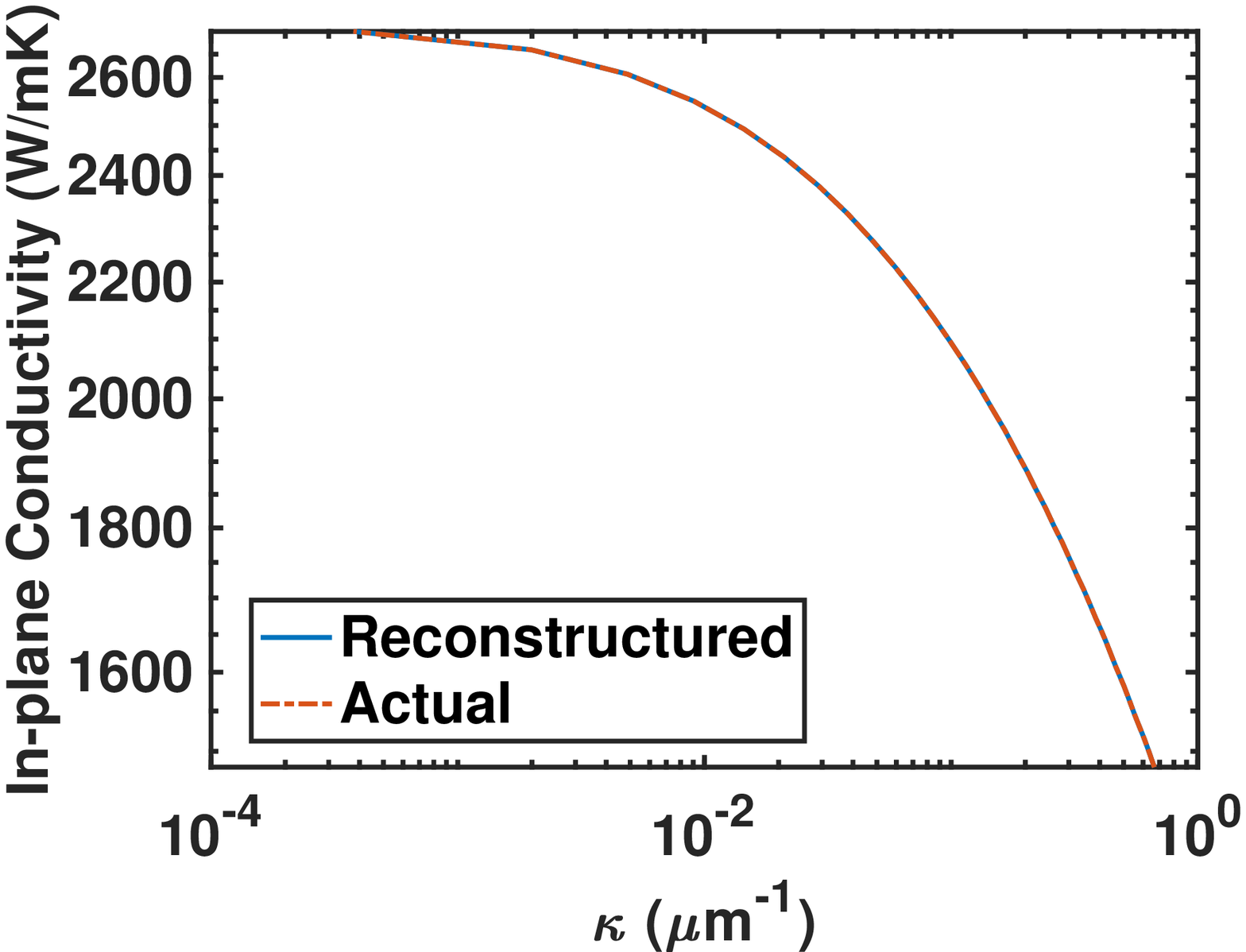}}
    \end{subfigure}
              \centering
    \begin{subfigure}[]{
        \includegraphics[scale=0.4]{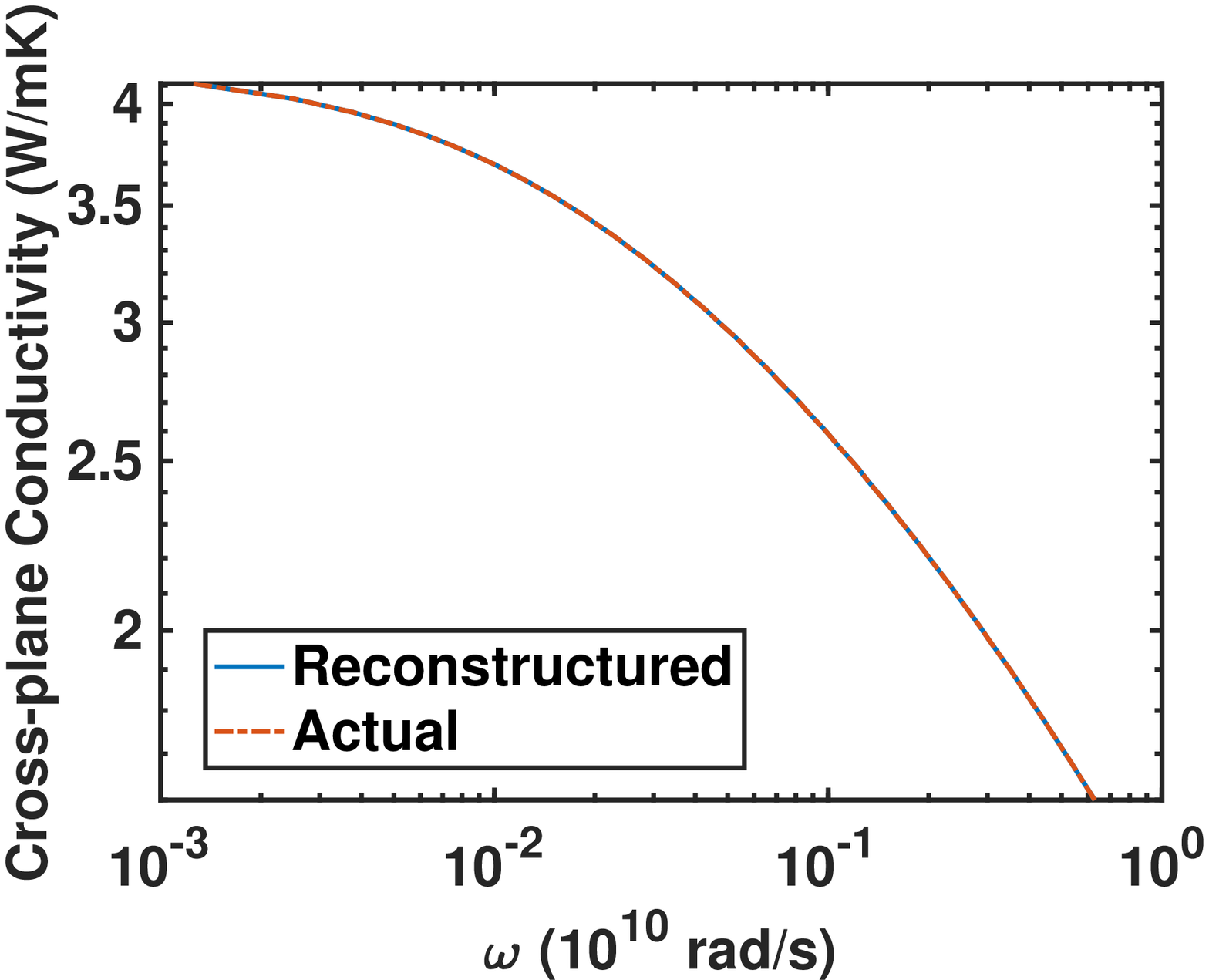}}
    \end{subfigure}
    \caption{(a) Temperature versus time plot of graphite with the same initial heating size as Fig. \ref{fig:aniso_case}(c). The temperature fall is much sharper than in Fig.\ref{fig:aniso_case}(c) due to the much higher in-plane thermal conductivity of graphite. (b) Transformed $\bar{\Delta T}(\kappa,\omega)$ from the temperature data in (c) using Eq. \ref{eq:hankel_fourier}. The transformed data is plotted in log scale of the absolute value as the values are typically complex. (c,d) Reconstructed in-plane (c) and cross-plane (d) thermal conductivity as a function of spatial wavevector $\kappa$ (c) and time domain frequency $\omega$ (d) versus the reference spectrum. The reference spectrums are generated using phonon mean-free-path distributions of graphite \cite{minnich_phonon_2015} convoluted with experimental suppression functions in spatial \cite{ding_radial_2014} and time domains \cite{yang_heating-frequency-dependent_2015}.  }\label{fig:graphite}
\end{figure}

The next example involves the use of the same method but assume a size-dependent thermal conductivity. This means that $k_{r,z}$ is a function of the transformed space parameters $(\kappa,\omega)$. Here, we choose graphite as a test case. This is because graphite is highly anisotropic, making the thermal transport of great interest. Furthermore, graphite has a in plane phonon mean-free-path (MFP) on that range of micrometers at room temperature while a small phonon MFP of hundred nanometers in cross plane (or the c-axis). Recent effort to resolve c-axis MFP requires careful measurement of exfoliated samples of different thicknesses\cite{zhang_temperature_2015} . In order to obtain the dependence of $k_{r,z}$ on $(\kappa,\omega)$, we need to obtain the radial and time domain suppression function and the phonon MFP distribution of graphite in the radial and cross plane direction.  Here, we took phonon MFP data from Ref.\cite{minnich_phonon_2015}  and suppression function in radial \cite{ding_radial_2014} and frequency domain \cite{yang_heating-frequency-dependent_2015}. We assume that in-plane thermal conductivity is purely suppressed radially and that cross-plane thermal conductivity is suppressed purely in the time domain.

Figure \ref{fig:graphite}(a) shows the temperature versus time data of graphite with the same initial heating condition as the multilayer example (\ref{fig:aniso_case}(a)). One can see that graphite has a much faster decay time compared to multilayer case in Fig. \ref{fig:aniso_temp}(b) due to the high thermal conductivity. The data can be transformed in the same manner, resulting in a distribution in the phase space shown in \ref{fig:graphite}(b). Now, the system of equations is not going to be the case where all values $k_{r,z}$ are fixed. $k_{r,z}$ will be different for each set of $(\kappa,\omega)$. So we only have two equations for two unknowns by using the fact that $\bar\Delta T(\kappa,\omega)=\bar\Delta T(\kappa,\omega)^{\dagger}$. Solving $\bar\Delta T(\kappa,\omega)=\bar\Delta T(\kappa,\omega)^{\dagger}$ for each pair of $(\kappa,\omega)$, we assume a smooth solution of $k_{r,z}(\kappa,\omega)$ and obtain stable solutions for $k_{r,z}(\kappa,\omega)$. Here, we choose $N_k=50$ and $N_{\omega}=500$ in order to show a smooth size-dependent spectrum for $k_{r,z}(\kappa,\omega)$.

Figures \ref{fig:graphite}(c) and (d) shows the reconstructed $k_{r}(\kappa)$ and $k_{z}(\omega)$ respectively. We assume independence of $k_{r}$ from $\omega$ and $k_{z}(\kappa)$ from $\omega$ such that the in-plane thermal conductivity only depends on radial heating size and the cross-plane thermal conductivity only depends on cross-plane heating size due to frequency dependent penetration depth. The reconstruction is almost in perfect agreement with the input size-dependent thermal conductivities, showing the potential of this method to accurately retrieve size-dependent thermal conductivity for various materials. 

\section{Discussion and Conclusion}


One proposed experimental method to realize our scheme used to retrieve spatial temporal temperature data which can be captured by regular CCD cameras. Our method can be adapted to any optical transient methods such as time-domain thermoreflectance \cite{jiang_tutorial:_2018} where optical reflectance is used. Also, IR detectors in laser flash methods can also be used \cite{zhao_measurement_2016}, albeit at slower response time and poorer spatial resolution. Alternatively, nanoscale scanning or imaging methods \cite{gomes_temperature_2005, mecklenburg_nanoscale_2015,laraoui_imaging_2015,kilbane_far-field_2016} can be used to retrieve temperature data point-by-point and with much better spatial resolution. This requires potentially longer data collection. Nevertheless, no heating size or sample size is required to be changed. 

If materials have in-plane anisotropy such as black phosphorous, regular spatial decomposition in Cartesian coordinates can be performed to obtain $\bar{\Delta T}$ in the transformed coordinates. More complicated optical excitation patterns such as transient grating can also be used and transformed according to the symmetry of the system. Essentially, our method opens up the concept of data multiplexing into thermal transport measurements. Now, instead of sending and retrieving one frequency at a time, we can retrieve information encoded over various frequencies by sampling and decode all at once.

Last but not least, our method offers a snapshot technique to gather size-dependent thermal conductivity information with the least experimental effort possible. Using methods in Machine Learning, we can potentially reduce the amount of spatial temporal data required to obtain thermal transport parameters \cite{zhang_machine_2018,xie_phonon_2018}. Furthermore, novel effects such as coherent phonons \cite{luckyanova_coherent_2012,ravichandran_crossover_2014} and hydrodynamic transport \cite{lee_hydrodynamic_2015}  can potentially be directly observed from one experimental setting, provided the correct features in our transformed spectrum are identified. In short, our method can provide direct diagnosis for systematic design of nanostructured materials to modify the overall thermal properties.

%


\end{document}